\documentstyle[twocolumn,aps,pra,epsf]{revtex}
\begin{document}
\draft
\title{Multiparty multilevel Greenberger-Horne-Zeilinger states}
\author{Ad\'{a}n Cabello\thanks{Electronic address:
adan@cica.es, fite1z1@sis.ucm.es}}
\address{Departamento de F\'{\i}sica Aplicada II,
Universidad de Sevilla, 41012 Sevilla, Spain}
\date{\today}
\maketitle

\begin{abstract}
The proof of Bell's theorem without inequalities by
Greenberger, Horne, and Zeilinger (GHZ) is extended
to multiparticle multilevel systems.
The proposed procedure generalizes previous partial results
and provides an operational characterization of
the so-called GHZ states for multiparticle multilevel systems.
\end{abstract}
\narrowtext

\section{Introduction}
Greenberger, Horne, and Zeilinger (GHZ) \cite{GHZ89,GHSZ90} show that the
quantum predictions for an individual system
composed of three or more particles prepared in a specific entangled
state (henceforth called a ``GHZ state'') cannot be reproduced by any local
hidden-variables model based on the definition of ``elements
of reality'' proposed by Einstein, Podolsky, and Rosen (EPR) \cite{EPR35}.
The proof was originally developed for four spin-$\frac{1}{2}$
particles \cite{GHZ89},
later simplified to three spin-$\frac{1}{2}$ particles
\cite{GHSZ90,Mermin90}, and has been
recently verified experimentally
\cite{BPDWZ99,PBDWZ00}.
On the other hand, the GHZ proof has been extended
to $n$ spin-$\frac{1}{2}$ particles \cite{PRC91,Cereceda95}, and to
three {\em pairs} of spin-$\frac{1}{2}$ particles \cite{Cabello99}.
In addition, GHZ-like proofs for particular states of $n$ spin-$s$ particles
have been investigated \cite{ZK99}.
So far, however, no generalization of the
GHZ proof for multiparty multilevel systems has been considered.
This extension would be of interest since it would lead to a definition
of GHZ states for multiparticle multilevel systems.
Particularly because the classification of the pure states
of single copies
of multiparticle multilevel systems becomes highly complicated beyond
three qubits \cite{AACJLT00,CDV}.

The structure of this paper is as follows: in Sec.~\ref{sec:II}, a
GHZ-like proof for three three-level systems is proposed.
In Sec.~\ref{sec:III}, the GHZ proof is extended to three $m$-level
systems. Additionally, in Sec.~\ref{sec:IV},
a proof of the Kochen-Specker theorem
\cite{KS1} based on the GHZ-like proof for the case of $m$
being an even number is presented.
In Sec.~\ref{sec:V}, the GHZ proof is generalized to $n$ particles of $m$ levels,
and to $n$ particles with varying number of levels (all with
the same parity). The main point of this paper is not the
generalization itself but to illustrate the basic ingredients of {\em any}
GHZ-like proof, so that we can provide a natural definition of what could
be called a ``GHZ state'' in the context of multiparty multilevel systems.
This definition is presented in Sec.~\ref{sec:VI}.

\section{GHZ-like proof for three three-level systems}
\label{sec:II}
The common scenario to any GHZ-like proof is the following:
A system composed of three (or $N \ge 3$) particles is initially
prepared in a specific pure entangled state (a GHZ state).
Each particle moves away to a distant space-time region where an observer
measures either $A_i$ or $B_i$, where $i$
denotes particle $i$. Local measurements on particle $i$
are assumed to be spacelike separated from local measurements
on the other particles.
For certain combinations of measurements, if all the
observers except observer $N$ share their results,
then they can predict with certainty the result
of measuring $A_N$ (or $B_N$) on particle $N$.
Therefore, adopting the
EPR criterion of elements of reality \cite{EPR35},
there must be an element of reality (a value)
corresponding to $A_N$ ($B_N$).
Similar reasonings lead us to conclude that
all the one-particle observables $A_i$ and $B_i$
must have predefined values.
The proof concludes by showing that
one cannot assign values to all these
one-particle observables in a way consistent with all
the quantum predictions.

In order to construct a GHZ proof
for the case of three three-level subsystems
(for instance, three spin-1 particles),
we must look for two one-particle maximal operators for
each quantum three-level subsystem (or ``qutrit'') $i$, $A_i$ and $B_i$,
such that they anticommute, that is
\begin{equation}
{A_i} {B_i}=-{B_i} {A_i}.
\end{equation}
For instance \cite{com},
\begin{eqnarray}
A_i & = & \left( {\matrix{1&0&0\cr
0&0&0\cr
0&0&{-1}\cr
}} \right),
\label{opA} \\
B_i & = & \left( {\matrix{0&0&1\cr
0&0&0\cr
1&0&0\cr
}} \right).
\label{opB}
\end{eqnarray}
Then, automatically, the four operators defined as
\begin{eqnarray}
A_{1}B_{2}B_{3}=A_{1}\otimes B_{2}\otimes B_{3},
\label{op1}\\
B_{1}A_{2}B_{3}=B_{1}\otimes A_{2}\otimes B_{3},
\label{op2}\\
B_{1}B_{2}A_{3}=B_{1}\otimes B_{2}\otimes A_{3},
\label{op3}\\
A_{1}A_{2}A_{3}=A_{1}\otimes A_{2}\otimes A_{3}.
\label{op4}
\end{eqnarray}
are mutually commutative, and therefore possess a set
of common eigenvectors.
The eigenvalues of these four operators are $-1$ (four of them), $0$
(nineteen of them),
and $1$ (four of them). In addition, the product
\begin{equation}
C = A_{1}B_{2}B_{3} \times B_{1}A_{2}B_{3} \times B_{1}B_{2}A_{3}
\times A_{1}A_{2}A_{3}
\label{prod}
\end{equation}
is a negative operator; its eigenvalues are either $-1$
(eight of them) or $0$ (nineteen). Suppose we chose a common eigenvector
$\left| {\mu } \right\rangle$ of the four
operators such that
\begin{eqnarray}
A_{1}B_{2}B_{3}\left| {\mu } \right\rangle & = &
\left| {\mu } \right\rangle,
\label{mu1} \\
B_{1}A_{2}B_{3}\left| {\mu } \right\rangle & = &
\left| {\mu } \right\rangle,
\label{mu2} \\
B_{1}B_{2}A_{3}\left| {\mu } \right\rangle & = &
\left| {\mu } \right\rangle,
\label{mu3} \\
A_{1}A_{2}A_{3}\left| {\mu } \right\rangle & = &
- \left| {\mu } \right\rangle.
\label{mu4}
\end{eqnarray}
This eigenvector is
\begin{equation}
\left| {\mu } \right\rangle = {1 \over 2}
\left( {\left| 11\bar{1} \right\rangle +
\left| 1\bar{1}1 \right\rangle +
\left| \bar{1}11 \right\rangle -
\left| \bar{1}\bar{1}\bar{1}\right\rangle
} \right),
\label{statemu}
\end{equation}
where
\begin{equation}
\left| 1 \right\rangle =
\left( {\matrix{1\cr
0\cr
0\cr
}} \right),\,\,\,\,
\left| 0 \right\rangle =
\left( {\matrix{0\cr
1\cr
0\cr
}} \right),\,\,\,\,
\left| \bar{1} \right\rangle =
\left( {\matrix{0\cr
0\cr
1\cr
}} \right).
\end{equation}
Now consider three observers, each having access to one particle.
On particle $i$ the corresponding observer measures
either $A_{i}$ or $B_{i}$ without disturbing the other particles.
The results of these measurements will be called
$a_{i}$ or $b_{i}$, respectively.
Since these results must satisfy the same functional
relations satisfied by the corresponding operator, then, from
Eq.~(\ref{mu1}), we can predict that, if $A_{1}$, $B_{2}$, and $B_{3}$
are measured, their results must satisfy
\begin{equation}
a_{1}b_{2}b_{3} = 1.
\label{nu1}
\end{equation}
Analogously, from Eqs.~(\ref{mu2})-(\ref{mu4}),
the results of other possible measurements must satisfy
\begin{eqnarray}
b_{1}a_{2}b_{3} & = & 1,
\label{nu2} \\
b_{1}b_{2}a_{3} & = & 1,
\label{nu3} \\
a_{1}a_{2}a_{3} & = & -1.
\label{nu4}
\end{eqnarray}
We can associate each of the eigenvalues $a_{i}$ and $b_{i}$
to an EPR element of reality \cite{EPR35} of particle $i$,
initially hidden in the original state of the system, but
``revealed'' by performing measurements on the other two
distant particles. For example, if the observers on particles 1 and 2
measure, respectively, $A_{1}$
and $B_{2}$, and their results are both $1$, then, sharing their results
and using Eq.~(\ref{nu1}), they can predict with certainty that the result of
measuring $B_3$ will be $1$.
Since arriving to this conclusion does not
require any real interaction with particle 3,
then, according to EPR, particle 3
has the value 1 for $B_3$, so we can assign
the value $1$ to the observable $B_{3}$.
Alternatively, since a different
measurement on particles 1 and 2 (for instance,
by measuring $B_{1}$ instead of $A_{1}$)
allows the observers of particles 1
and 2 to predict with
certainty, and without interacting with particle 3, the result of $A_3$
---using Eq.~(\ref{nu3})---,
then we suppose that this result was somehow predetermined.
Such predictions with certainty
would lead us to assign values
to the six observables $A_{1}$, $B_{1}$, $A_{2}$, $B_{2}$,
$A_{3}$, and $B_{3}$.
However, such assignment cannot be consistent with the rules of
quantum mechanics because
the four equations (\ref{nu1})-(\ref{nu4}) cannot
be satisfied simultaneously, since the product of their left-hand sides is
a positive number (because each value appears twice), whereas
the product of the right-hand sides is $-1$.
Therefore, the values of these observables
cannot be predefined as we assumed.
Note that we can also develop a similar reasoning if we choose any other
common eigenvector of the four operators so that the product of the
corresponding eigenvalues is negative.

\section{GHZ-like proof for three $m$-level systems}
\label{sec:III}
The method used in the previous section can be easily extended to any system
composed of three spin-$s$ (or $m$-level, with $m=2 s+1$) systems.
We will distinguish between the case of $m$ being an odd number
and the case of $m$ being an even number.

If $m$ is an odd number, we can choose the following one-particle
anticommutative operators for each subsystem $i$:
\begin{eqnarray}
A_i & = & \left( {\matrix{s&{}&{}&{}&{}&{}&{}&{}&{}\cr
{}&{s-1}&{}&{}&{}&{}&{}&{}&{}\cr
{}&{}&{...}&{}&{}&{}&{}&{}&{}\cr
{}&{}&{}&1&{}&{}&{}&{}&{}\cr
{}&{}&{}&{}&0&{}&{}&{}&{}\cr
{}&{}&{}&{}&{}&{-1}&{}&{}&{}\cr
{}&{}&{}&{}&{}&{}&{...}&{}&{}\cr
{}&{}&{}&{}&{}&{}&{}&{-s+1}&{}\cr
{}&{}&{}&{}&{}&{}&{}&{}&{-s}\cr
}} \right),
\label{Aodd} \\
B_i & = & \left( {\matrix{{}&{}&{}&{}&{}&{}&{}&{}&s\cr
{}&{}&{}&{}&{}&{}&{}&{s-1}&{}\cr
{}&{}&{}&{}&{}&{}&{...}&{}&{}\cr
{}&{}&{}&{}&{}&1&{}&{}&{}\cr
{}&{}&{}&{}&0&{}&{}&{}&{}\cr
{}&{}&{}&1&{}&{}&{}&{}&{}\cr
{}&{}&{...}&{}&{}&{}&{}&{}&{}\cr
{}&{s-1}&{}&{}&{}&{}&{}&{}&{}\cr
s&{}&{}&{}&{}&{}&{}&{}&{}\cr
}} \right).
\label{Bodd}
\end{eqnarray}
The other entries in the matrices are assumed to be zeroes.
The argument of nonlocality is almost identical to that in Sec.~\ref{sec:II}.
The only difference is that in this case the eigenvalues of the four operators
(\ref{op1})-(\ref{op4}) are positive ($[(2s+1)^3-k]/2$ of them, being
$k=12 s^2+6 s+1$), zero ($k$ of them), and negative
($[(2s+1)^3-k]/2$ of them), and the eigenvalues of product (\ref{prod})
are negative ($[2s+1]^3-k$ of them) and zero ($k$ of them).

If $m$ is an even number, we can choose the following one-particle
anticommutative operators:
\begin{eqnarray}
A_i & = & \left( {\matrix{s&{}&{}&{}&{}&{}&{}&{}\cr
{}&{s-1}&{}&{}&{}&{}&{}&{}\cr
{}&{}&{...}&{}&{}&{}&{}&{}\cr
{}&{}&{}&1&{}&{}&{}&{}\cr
{}&{}&{}&{}&{-1}&{}&{}&{}\cr
{}&{}&{}&{}&{}&{...}&{}&{}\cr
{}&{}&{}&{}&{}&{}&{-s+1}&{}\cr
{}&{}&{}&{}&{}&{}&{}&{-s}\cr
}} \right),
\label{AS3} \\
B_i & = & \left( {\matrix{{}&{}&{}&{}&{}&{}&{}&s\cr
{}&{}&{}&{}&{}&{}&{s-1}&{}\cr
{}&{}&{}&{}&{}&{...}&{}&{}\cr
{}&{}&{}&{}&1&{}&{}&{}\cr
{}&{}&{}&1&{}&{}&{}&{}\cr
{}&{}&{...}&{}&{}&{}&{}&{}\cr
{}&{s-1}&{}&{}&{}&{}&{}&{}\cr
s&{}&{}&{}&{}&{}&{}&{}\cr
}} \right),
\label{BS3}
\end{eqnarray}
and develop a similar argument. The big difference in the case
in which $m$ is an even number
is that the one-particle operators (\ref{AS3}) and (\ref{BS3}) have no zero
eigenvalues, so the product of the four operators (\ref{prod}) is a definite
negative operator (i.e., all its eigenvalues are negative). Thus, {\em every}
common eigenvector of the four operators will allow us to develop a GHZ-like
argument.

\section{Kochen-Specker proof for three $m$-level systems, with $m$
being an even number}
\label{sec:IV}
Indeed, the last result of the previous section allows us to develop a
``multiplicative'' proof of the Kochen-Specker theorem \cite{KS1}
for a three $m$-level system, being $m$ an even number, that
generalizes those multiplicative proofs proposed by Mermin
for three spin-$\frac{1}{2}$ particles
\cite{KS2} or by Cabello for three pairs of spin-$\frac{1}{2}$ particles
\cite{Cabello99}.
As seen above, in the case of $m$ being an even number,
a GHZ-like proof could be developed starting
from any common
eigenvector of the four operators (\ref{op1})-(\ref{op4}).
Therefore, the argument can be rearranged as a state-independent
proof of the Kochen-Specker theorem in an $m^3$-dimensional Hilbert space
(with $m$ being an even number) just with
the inclusion of these four operators.
The resulting proof of the Kochen-Specker theorem is summarized in Fig. 1,
which contains ten operators: the four operators (\ref{op1})-(\ref{op4})
acting on the whole system, and the six one-particle operators $A_i$ and
$B_i$. The four operators on
each of the five straight lines are mutually commutative. As stated above,
the product of the four operators on the horizontal line is
a definite negative operator and,
as can be easily verified, the product
of the four operators on each of the other lines is one and the
same definite positive operator.
It can be easily checked that it is impossible
to ascribe one of their eigenvalues to each of the ten operators,
satisfying the same functional relations that are satisfied by
the corresponding operators.

\section{GHZ-like proof for $n$ $m$-level systems}
\label{sec:V}
Let us extend the GHZ proof to the case of a system with $n$
subsystems, all of them with $m$ levels.
It is convenient to distinguish between the case in
which $n$ is an odd number and the case in which $n$ is an even number.

In case of $n$ being an odd number, the proofs
(for $m$ being an odd number, or for $m$ being an even number) will be similar
to the proofs in Sec.~\ref{sec:III}:
We will use the same one-particle operators $A_i$
and $B_i$ (now with $i$ from 1 to $n$), and we will construct four
$n$-particle operators of the form $O_{1} \otimes O_{2} \otimes ... \otimes
O_{n}$, where $O_i$ is $A_i$ or $B_i$. These $n$-particle operators must
satisfy the following requirements \cite{comC}: (i) In order to commute,
the $n$-particle operators must contain a number of operators of the
$A_i$ kind with the same parity in all of them (thus the parity of the
operators $B_i$ will also be the same). (ii) In order to avoid the
product of the four $n$-particle operators having positive eigenvalues, one of
the $n$-particle operators must have a different number (but with the same
parity) of operators of the
kind $A_i$ than the other three. (iii) In order to obtain a GHZ-like algebraic
(parity) contradiction, each one-particle operator must be used in the
construction of two $n$-particle operators. (iv) In order to obtain a
nontrivial proof (in the sense that all particles are required for the
contradiction), all one-particle operators must be used in the definition
of the $n$-particle operators.
For instance, for $n=5$, the following four operators allow us to develop a
GHZ-like proof in a similar way as in Sec.~\ref{sec:III}:
\begin{eqnarray}
A_{1} \otimes B_{2} \otimes B_{3} \otimes B_{4} \otimes B_{5}, \\
A_{1} \otimes A_{2} \otimes A_{3} \otimes B_{4} \otimes B_{5}, \\
B_{1} \otimes B_{2} \otimes A_{3} \otimes A_{4} \otimes A_{5}, \\
B_{1} \otimes A_{2} \otimes B_{3} \otimes A_{4} \otimes A_{5}.
\end{eqnarray}
The one-particle operators $A_i$ and $B_i$ can be, respectively,
those of Eqs.~(\ref{Aodd}) and (\ref{Bodd}), if $m$ is an odd number,
or those of Eqs.~(\ref{AS3})
and ({\ref{BS3}), if $m$ is an even number.

The case of $n$ particles with $n$ being an even number
is more complicated since, as can be
easily checked, the requirements (i)-(iv) cannot be satisfied. A simple trick
for developing a proof in this case is as follows:
Start with a GHZ proof for $n-1$ particles, then
construct four $n$-particle operators just by adding
the same one-particle operator to the four
$(n-1)$-particle operators
(i.e., by making their tensor product with
the same one-particle operator). For instance, in order to construct a
proof for $n=4$, let us start with the four three-particle operators
given by Eqs.~(\ref{op1})-(\ref{op4}), and add the one-particle operator
$B_4$. This leads to the following four-particle operators:
\begin{eqnarray}
A_{1} \otimes B_{2} \otimes B_{3} \otimes B_{4},
\label{o1} \\
B_{1} \otimes A_{2} \otimes B_{3} \otimes B_{4},
\label{o2} \\
B_{1} \otimes B_{2} \otimes A_{3} \otimes B_{4},
\label{o3} \\
A_{1} \otimes A_{2} \otimes A_{3} \otimes B_{4}.
\label{o4}
\end{eqnarray}
Note that this set of operators does not satisfy (iv).
In order to fulfill (iv), we can add one new four-particle operator
ending on $A_{4}$ and containing a number of $A_i$ and $B_i$ with the
same parity as the operators (\ref{o1})-(\ref{o4}).
For instance,
\begin{equation}
B_{1} \otimes B_{2} \otimes B_{3} \otimes A_{4}.
\label{five}
\end{equation}
Note that then the product of the operators (\ref{o1})-(\ref{five})
would contain
positive eigenvalues. However, in order to avoid this, we can
consider the product of six operators, being the new one the same
defined in (\ref{five}). The rest of the proof is as in Sec.~\ref{sec:III}.

The same method can be applied to any system composed of $n$ parts
with, respectively, $m_1$, $m_2$, \dots, $m_n$ levels.
The only restriction is
that $m_1$, $m_2$, \dots, $m_n$ must have the same parity.

On the other hand, if all the $n$-particle operators have the same eigenvalues
and none of them is zero, then one can develop a Kochen-Specker multiplicative
proof for the corresponding Hilbert space in a similar way
to that of Sec.~\ref{sec:IV}.

\section{What is a GHZ state?}
\label{sec:VI}
The aim of this work is not only to show how to
develop a GHZ-like proof for multiparty multilevel quantum systems, but also
to illustrate the basic ingredients of any GHZ-like proof, and provide
a natural definition of GHZ states in this context.
This is of interest since the classification of pure states of
single copies of composite
systems becomes more difficult
as we increase the number of parts or the number of levels. In fact,
this classification is highly difficult beyond three-particle
two-level systems \cite{AACJLT00,CDV}.
Therefore, the question of what is a GHZ state for a multiparty multilevel
quantum system is
not trivial. According to the proofs presented in this paper, a natural
definition of GHZ states for multiparticle multilevel systems is any for
which: (I) for every subsystem there are two one-particle (maximal or not)
anticommutative operators $A_i$ and $B_i$, such that (II) the state
of the system is an
eigenvector of a set of four (in the case of subsystems with an even number
of levels) or five (in the case of subsystems with an odd number of levels)
$n$-particle operators constructed as a tensor product of these one-particle
operators, and such that (III) the product of the corresponding eigenvalues
leads to an algebraic (parity) contradiction if one assumes EPR elements of
reality. For instance, these criteria allow us to say that the three two-level
state, given by \cite{CDV}
\begin{equation}
\left| {W} \right\rangle =
{1 \over {\sqrt{3}}}
\left( {\left| 001 \right\rangle +
\left| 010 \right\rangle +
\left| 100 \right\rangle
} \right),
\label{stateW}
\end{equation}
where
\begin{equation}
\left| 0 \right\rangle =
\left( {\matrix{1\cr
0\cr
}} \right),\,\,\,\,
\left| 1 \right\rangle =
\left( {\matrix{0\cr
1\cr
}} \right),
\end{equation}
is not a GHZ state, since it is not a common eigenvector of four
commuting operators of the form $O_{1} \otimes O_{2} \otimes O_{3}$,
although
it is a ``maximally entangled'' state in the sense described in \cite{CDV}.
In addition, the proofs
presented in this work provide a constructive method to generate GHZ states
for any multiparty multilevel (all with the same parity) system.

\section*{Acknowledgements}
The author thanks R. Tarrach
for suggesting this problem, J. Calsamiglia, J. L. Cereceda,
and C. Serra
for useful comments, M. \.Zukowski for bringing Ref.~\cite{ZK99} to
his attention, and the
organizers of the Sixth Benasque Center for Science for support.

\begin{figure}
\epsfxsize=8.6cm
\epsffile{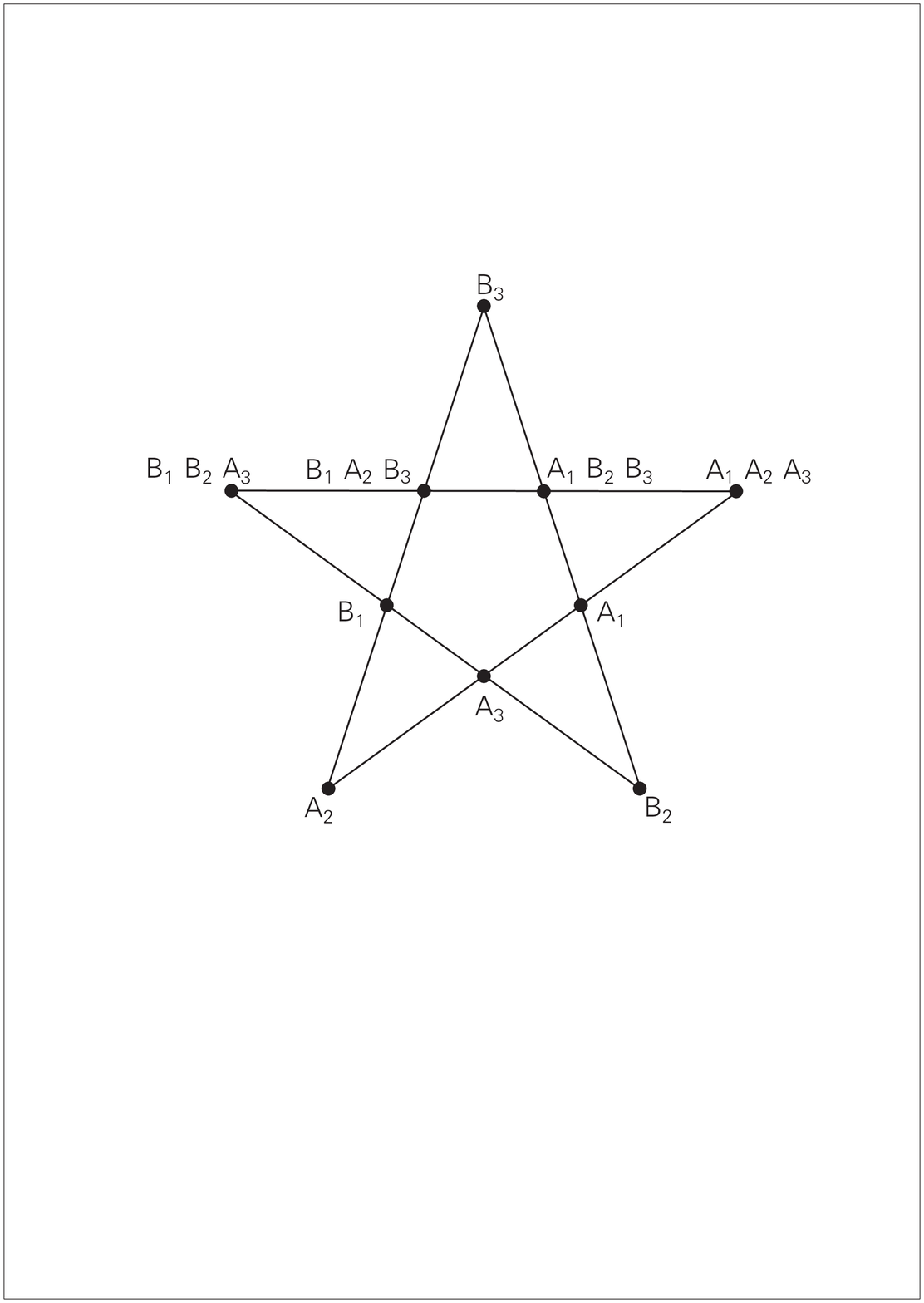}
\end{figure}
\noindent FIG. 1:
{\small Each dot represents an observable.
The ten observables provide a proof of
the Kochen-Specker theorem
in a Hilbert space of dimension $m^3$, with $m$ being an even
number.
The four observables on each line are mutually compatible
and the product of their results must be positive, except for the
horizontal line, where the product must be negative.}

\begin{thebibliography}{99}
\bibitem{GHZ89} D.M. Greenberger, M.A. Horne,
and A. Zeilinger, in {\em Bell's
Theorem, Quantum Theory, and Conceptions of
the Universe}, edited by M. Kafatos
(Kluwer, Dordrecht, 1989), p. 69.
\bibitem{GHSZ90} D.M. Greenberger, M.A. Horne, A. Shimony, and
A. Zeilinger, Am. J. Phys. {\bf 58}, 1131 (1990).
\bibitem{EPR35} A. Einstein, B. Podolsky, and N. Rosen,
Phys. Rev. {\bf 47}, 777 (1935).
\bibitem{Mermin90} N.D. Mermin, Phys. Today {\bf 43}(6), 9 (1990);
Am. J. Phys. {\bf 58}, 731 (1990).
\bibitem{BPDWZ99} D. Bouwmeester, J. Pan, M. Daniell, H. Weinfurter,
and A. Zeilinger,
Phys. Rev. Lett. {\bf 82}, 1345 (1999).
\bibitem{PBDWZ00} J. Pan, D. Bouwmeester, M. Daniell, H. Weinfurter,
and A. Zeilinger,
Nature (London) {\bf 403}, 515 (2000).
\bibitem{PRC91} C. Pagonis, M.L.G. Redhead, and R.K. Clifton,
Phys. Lett. A {\bf 155}, 441 (1991).
\bibitem{Cereceda95} J.L. Cereceda,
Found. Phys. {\bf 25}, 925 (1995).
\bibitem{Cabello99} A. Cabello,
Phys. Rev. A {\bf 60}, 877 (1999).
\bibitem{ZK99} M. \.Zukowski and D. Kaszlikowski,
Phys. Rev. A {\bf 59}, 3200 (1999).
\bibitem{AACJLT00} A. Ac\'{\i}n, A. Andrianov, L. Costa, E. Jan\'{e},
J.I. Latorre, and R. Tarrach,
Phys. Rev. Lett. {\bf 85}, 1560 (2000).
\bibitem{CDV} W. D\"{u}r, G. Vidal, and J.I. Cirac,
Los Alamos e-print archive quant-ph/0005115.
\bibitem{KS1} S. Kochen and E.P. Specker,
J. Math. Mech. {\bf 17}, 59 (1967).
\bibitem{com} We can develop similar GHZ-like arguments starting from
one-particle maximal operators with different eigenvalues
other than those
chosen in this paper, or even starting from non-maximal operators.
However, the choice in this paper (maximal operators with eigenvalues from
$-s$ to $s$) leads to a natural generalization of the original GHZ proof with
spin-$\frac{1}{2}$ systems.
After this work was completed it has been proved that,
for three spin-$s$ particles,
any one-particle operator can always be reduced to a direct sum of
two two-level anticommuting one-particle operators,
J. Savinien, J. Taron, and R. Tarrach,
Los Alamos e-print archive quant-ph/0007069.
\bibitem{KS2} N.D. Mermin, Phys. Rev. Lett. {\bf 65}, 3373 (1990);
Rev. Mod. Phys. {\bf 65}, 803 (1993).
\bibitem{comC} A similar set of requirements for a GHZ-like contradiction in
the case of $n$ spin-$\frac{1}{2}$ particles is proposed in Ref.~\cite{PRC91}.
\end{thebibliography}
\end{document}